\documentclass[a4paper,11pt]{article} 

\usepackage[a4paper,top=1.2cm,bottom=1.8cm,left=1.4cm,right=1.4cm,marginparwidth=1.75cm]{geometry}

\usepackage[T1]{fontenc}
\usepackage{lmodern}
\usepackage[utf8x]{inputenc}
\usepackage[english]{babel}
\usepackage[normalem]{ulem} %
\usepackage{stackengine}  %
\usepackage[table]{xcolor}    %

\usepackage{amsmath}
\usepackage{graphicx}
\usepackage{ulem}
\usepackage[colorinlistoftodos]{todonotes}
\usepackage{url}
\usepackage{xfrac}
\usepackage{cite}
\everypar{\looseness=-1}

\usepackage{wrapfig}
\usepackage{graphicx} 
\usepackage{float}

\usepackage[colorlinks=true, allcolors=blue]{hyperref}

\title{Future Opportunities with Lepton-Hadron 
Collisions\footnote{Submission to the 2025 European Particle Physics Strategy Update.}}

\author{Contact persons:
  Allen Caldwell\rlap,$^1$
  Silvia Dalla Torre\rlap,$^2$
  Rolf Ent\rlap,$^3$ \\
  Aharon Levy\rlap,$^4$
  Paul Newman\rlap,$^5$
  Fred Olness$^6$ and
  Juan Rojo$^{7,8}$
  \\[2ex]   \small \it
${}^1$Max Planck Institute for Physics, Munich, Germany
\\   \small \it
${}^2$INFN - Sezione di Rrieste, I-34149, Trieste, Italy
\\   \small \it
${}^3$Thomas Jefferson National Accelerator Facility, Newport News, VA 23606, USA
\\     \small \it
${}^4${Raymond \& Beverly Sackler School of Physics \& Astronomy, Tel Aviv University, Tel Aviv, Israel}
\\   \small \it
${}^5$School of Physics and Astronomy, University of Birmingham, Birmingham, UK
 \\    \small \it
${}^6$Department of Physics, Southern Methodist University, Dallas, TX 75275 USA 
\\    \small \it
${}^7$Nikhef, Science Park 105, 1098 XG Amsterdam, The Netherlands 
\\    \small \it
${}^8$Department of Physics and Astronomy, Vrije Universiteit, NL-1081 HV Amsterdam, The Netherlands
 \vspace{0.5cm}
}

\begin{document}
\maketitle
\begin{abstract}

Deep Inelastic lepton-hadron Scattering (DIS) is a cornerstone of particle physics discovery and the precision measurement of the structure of matter. 
This document surveys the international
DIS landscape, exploring current and future opportunities to 
continue this rich heritage, leading to new understandings and 
enabling discoveries. 
Of immediate relevance to the future of the field in Europe, 
the Large Hadron electron
Collider (LHeC) offers an impactful bridge between 
the end of the HL-LHC and the beginning of the 
next CERN flagship project, both in terms of
technology
development and new scientific exploration from Higgs physics to 
the partonic structure of the proton. 

More generally, the 
facilities described here cover centre-of-mass energies
from a few GeV to multiple TeV and 
address a wide range of physics
topics, with unique sensitivity to
Quantum Chromodynamics and hadron structure
at their core. In addition to their stand-alone importance,
these topics enhance the precision measurement and 
new physics search programmes 
at hadron-hadron colliders. 

The very high luminosity fixed-target 
CEBAF programme that is in progress
at Jefferson Laboratory
probes nucleon and light ion structure at large $x$
in novel ways, while high energy neutrino DIS is being 
enabled at the FASER and SND@LHC experiments by the 
intense LHC 
beams;
both have exciting potential upgrade programmes. 
The Electron Ion Collider (EIC) is on course for deployment at Brookhaven 
in the early 2030s, 
and will provide lepton-nucleus and double-polarised lepton-proton/light-ion
collisions for the first time. 
Its science includes a 
3-dimensional
mapping of the internal structure and dynamics of hadrons, 
leading to
a thorough understanding of the mechanisms that generate 
proton mass and spin, whilst establishing accelerator
and detector technologies 
of direct relevance to 
next-generation facilities.
Adding the LHeC provides a Europe-based lepton-hadron frontier. 
The LHeC
extends DIS capabilities to include
a complementary Higgs, top and electroweak
programme to the HL-LHC,
together with
precise determinations of proton and nuclear 
structure 
in a kinematic range that improves HL-LHC
sensitivities.
In the 
longer term, plasma wakefield acceleration and the 
Future Circular Collider offer different possible pathways for major 
steps forward in centre-of-mass energy,
extending 
into a low parton momentum-fraction domain
where our present understanding fails and new
strong interaction 
discoveries are guaranteed. 

This review 
emerges from the `DIS and Related Subjects' conference series, 
which provides an annual focus for 
the diverse community of scientists involved in Deep Inelastic
Scattering, currently estimated to consist of around 
3000
experimental and theoretical 
particle, nuclear and accelerator physicists
worldwide.

\end{abstract}

\vspace*{2cm}

email contact: p.r.newman@bham.ac.uk

\pagenumbering{gobble} 
\newpage
\setcounter{page}{1} 
\pagenumbering{arabic}

\section{Executive Summary}
\label{intro}
\def\fnote#1{{\footnotesize \textcolor{blue}{[#1]}}}

Science is entering a new era in the investigation of matter, 
driven by a wealth of high precision measurements from a variety of experiments. When combined with revolutionary 
advances in theoretical tools and techniques, these precision analyses promise to answer fundamental open questions about the mass, spin, and structure of matter and, in parallel, to provide a solid foundation to enable future discoveries~\cite{Caldwell:2018wqk,Achenbach:2023pba,Amoroso:2022eow,Begel:2022kwp,FCC:2018byv,Feng:2022inv,Arguelles:2019xgp,AbdulKhalek:2021gbh,AbdulKhalek:2022hcn,Abir:2023fpo,LHeC:2020van,P5:2023wyd,NuPECC2024}.

Deep Inelastic Scattering (DIS) stands out as the archetypal process for high-precision measurement of matter. 
The ability of point-like leptons to probe the structure of hadronic targets yields unparalleled accuracy, crucial for the exploration of Quantum ChromoDynamics (QCD) and at the heart of our exploration of the Standard Model and new physics searches beyond~\cite{Achenbach:2023pba,Amoroso:2022eow,Begel:2022kwp}.

A new DIS energy frontier enabled by future $ep$ facilities allows access to discovery in extreme limits of QCD, provides direct sensitivity to key electroweak and Beyond the Standard Model (BSM) physics, and offers high sensitivity to Higgs couplings, including some of those most challenging to constrain at the HL-LHC. Combining $ee$, $ep$, and $pp$ results provides the best possible precision for  Higgs couplings~\cite{Arguelles:2019xgp}, which can yield incisive tests of the SM and constrain and/or reveal BSM physics.

Of the four fundamental forces of nature, the QCD theory of strong interactions is 
the most complex and enigmatic. QCD exhibits the dual characteristics of confinement of the quarks and gluons (at large distance scales) and asymptotic freedom (at short distance scales). These characteristics make QCD extremely challenging to formulate the disparate phenomena associated with both the
perturbative and the 
nonperturbative dynamics of strong interactions. 
Advances in QCD theory have opened new pathways to understanding how nature shapes the structure and dynamics of confined 3D hadrons, as well as how fundamental properties like mass and spin emerge from the intricate interactions of QCD.

This intricacy of QCD leads to fascinating non-linear collective effects. With current and future DIS facilities, we can explore QCD in extreme kinematic limits where we encounter a variety of exotic effects, including: quark-gluon plasma, collective phenomena, gluon saturation and recombination, jet quenching, and non-factorisable contributions. This endeavor is synergistic with a broad spectrum of research as diverse as dark matter searches and electroweak symmetry breaking.

Within the context of the QCD theory, the Parton Distribution Function (PDFs) framework is the primary computational tool available to describe the interactions of hadrons~\cite{Alekhin:2017kpj,Hou:2019efy,Bailey:2020ooq,NNPDF:2021njg,Ablat:2024muy,PDF4LHCWorkingGroup:2022cjn}. The PDFs encode the QCD dynamics and link the ``theoretical'' world of quarks and gluons with the ``experimentally measurable{}'' world of hadrons. Improved PDFs will result in more precise analyses of all 
current and future experimental results involving hadrons, 
providing insight into the underlying QCD dynamics and the resulting nuclear structure. 
Both the lepton and neutrino DIS processes are required for a complete hadronic flavor decomposition.

Beyond the (collinear) PDFs described above, we can now explore the 3-dimensional structure of hadrons. Transverse Momentum Dependent (TMD) PDFs encode nonperturbative information about the partonic transverse momentum and polarization  degrees of freedom, and 
naturally
lead to observable transverse momenta in the final 
state \cite{Angeles-Martinez:2015sea}. 
TMDs are essential for describing spin and azimuthal asymmetries, as well as non-inclusive processes (e.g., $p_T$ distributions of electroweak gauge bosons); the latter process directly impacts the $W$ boson mass measurement.

Similarly, the Generalized Parton Distributions (GPDs) describe the transverse position of a parton in a nucleon. Taken together, the TMDs and GPDs will play a prominent role in future precision experiments. As we expand high precision measurements to the edge of the perturbative domain, we can build connections to nonperturbative phenomena. Together with new theoretical tools (Lattice QCD, Higher-Order Calculations, Monte Carlo Generators, Machine Learning, Artificial Intelligence, etc.), we can better understand and describe complex nonperturbative phenomena. In particular, this will be a key goal of the upcoming EIC.

\makeatletter 
\newcommand\dotover{\leavevmode\cleaders\hb@xt@ 0.4em{\hss $\cdot$\hss}\hfill\kern\z@} 
\newcommand{\dotfrac}[2]{ \ooalign{$\genfrac{}{}{0pt}{0}{\mbox{\rm \small #1}}{\mbox{{\small \rm #2}}}$\cr\dotover\cr} } 
\makeatother

\def\fstrut{\vrule height 15pt depth 9pt width 0pt} 
\def\fstack#1#2{\fstrut \small \stackanchor{#1}{#2}}

\rowcolors{3}{gray!15}{white}
\begin{table*}[!ht]
\begin{centering}
\begin{tabular}{|l|l|l|l|l|l|l|}
\hline 
\textbf{Facility} & \textbf{Years} & \textbf{$E_{cm}$ } & \textbf{Luminosity} & \textbf{Ions} & \textbf{Polarisation} & \textbf{Status}\tabularnewline
 &  & (GeV)  & {\small{}($10^{33}/cm^{2}/s$) } & $^{*}${\footnotesize{}(depends on)} &  & \tabularnewline
\hline 
\hline
\fstrut {\dotfrac{JLab 11}{JLab 22}}  & \small \dotfrac{Running}{Late 2030's} & 4.5 --- 6.5 & $10^{2}$ --- $10^{6}$ & p $\to$ Pb  & e, p, \small \stackanchor{Light}{nuclei} & \small \dotfrac{Running}{Concept}\tabularnewline
\hline 
\fstrut {\small \dotfrac{FASER \phantom{xxxxx} }{FPF/AdvSND}}  &  {\small \  \dotfrac{Running}{2030's}}  & 30 --- 90 & 0.3 --- 10  & W, Ar & no & \small \dotfrac{Running}{Advanced}\tabularnewline
\hline 
~EIC \fstrut  & $>2034$  & 30 --- 140  & 1 --- 10 & p $\to$ U  & e,p,d,$^{3}$He & Approved\tabularnewline
\hline 
~EicC \fstrut  &  $>$\,Late 2030's  & 15---20  & 2 --- 3 & p $\to$ U  & e,p,d,$^{3}$He &  Concept\tabularnewline
\hline 
~LHeC \fstrut  &  $>$\,Late 2030's & 1200 & 24 & $^{*}$LHC  & e possible  & Advanced\tabularnewline
\hline 
\fstrut {\small \fstack{Plasma-based}{schemes} }  & \quad 2040's  & 530 --- 9000  & $10^{-5}$ --- $10^{-1}$ & $^{*}$SPS/LHC  & e possible  & Concept\tabularnewline
\hline 
~FCC-eh \fstrut  & $>2050$ & 3500  & 15 & $^{*}$FCC-hh  & e possible  & Concept\tabularnewline
\hline 
\end{tabular}\caption{Overview of selected DIS facilities.
}
\par\end{centering}
\label{tab:facilities} 
\end{table*}

To explore these phenomena, there are a variety of approved and proposed DIS facilities as detailed in  Table~\ref{tab:facilities}.
The rich DIS lepton-hadron heritage is continued by current facilities, including 
the Jefferson Lab CEBAF~\cite{Dudek:2012vr}  
neutrino DIS at CERN~\cite{Feng:2022inv,FLF:EPPSU20}, 
the forthcoming US-based %
EIC~\cite{AbdulKhalek:2021gbh,AbdulKhalek:2022hcn,Abir:2023fpo,EIC:EPPSU20,EIC:EPPSU20det,Acosta:2022ejc},
and the proposed Electron-ion collider in China (EicC)~\cite{Anderle:2021wcy}. 
The proposed LHeC~\cite{LHeC:2020van,Andre:2022xeh,lhec:whitepaper} programme is essential to provide a Europe-based DIS frontier. 
The LHeC complements the CERN flagship LHC and HL-LHC programme, 
can ensure the long-term success of the 
high energy 
physics collider research programme, 
and may serve as a bridge to a future flagship program, based for example on the
FCC~\cite{FCC:2018byv}.
Additionally,  Plasma wakefield acceleration techniques have the  potential to 
 explore the unknown at unprecedented lepton-hadron centre-of-mass energies \cite{Caldwell:2016cmw,ref:ALIVE}.

\goodbreak\null\vspace{12pt}
The impact of these facilities on physics is summarised in the list below.
\nobreak
\begin{itemize}
 \setlength\itemsep{0em}

    \item {\bf Fundamental Discovery:} 
    Enable precision tests of the Standard Model through stand-alone searches for BSM physics and sensitivity to deviations from predictions in the Higgs and Electroweak sectors that are complementary to those of $pp$~and $ee$~colliders.

    \item {\bf Fundamental Structure:} Understand how basic
    hadronic properties such as mass, spin, and 3D structure emerge from their quark and gluon constituents.

    \item {\bf  High Precision:}  Provide the precision QCD measurements required to study the SM and yield the most constraining limits on BSM physics in pp collisions while also enabling new insights into QCD through heavy ion collisions.

     \item {\bf Extreme Phenomena:} %
     Explore the extreme limits of QCD to understand and discover manifestations of non-linear collective effects. Validate the breakdown of linear QCD 
     and study the associated new high density
     and high energy phenomena.
     
    \item {\bf Interdisciplinary:} Contribute essential constraints for cosmic-ray air showers, neutrino astrophysics, and the AdS/CFT conjecture linking gravity-like theories to QCD-like strong coupling theories.

\end{itemize}

Precision experimental measurements will offer an unprecedented understanding of the strong force and its manifestations.
The DIS framework serves as a critical, high-precision tool for describing these interactions within the context of QCD theory. As such, it is the cornerstone for exploring uncharted kinematic domains, providing new insights into the mysteries of QCD, and paving the way for groundbreaking discoveries.

\section{Lepton-Hadron Facilities and their Scientific Objectives}
\label{facilities}

\subsection{The Electron Ion Collider}
\label{eic}

\noindent
{\bf Science Potential:}
The Electron-Ion Collider (EIC) is a powerful new facility to be built in the United States at the U.S. Department of Energy’s (DOE) Brookhaven National Laboratory (BNL) in partnership with Thomas Jefferson National Accelerator Facility (JLab). DOE, BNL, and JLab envision an EIC facility that is fully international in character.

Protons and neutrons, the building blocks of nuclear matter, constitute about 99.9 percent of the mass of all visible matter in the universe. These building blocks are themselves made up of quarks that are bound by gluons that also bind themselves. Thus, the interactions and structures are inextricably mixed, in sharp contrast with more familiar atoms and molecular systems. Indeed, the observed properties of nucleons and nuclei, such as their mass and spin, emerge from a complex, dynamical system governed by QCD,
the theory of strong interaction with quarks and gluons as the fundamental degrees of freedom. Consequently, the quark masses, generated via the Higgs mechanism, only account for a tiny fraction of the mass of a proton. Key science questions that the EIC will address are:
\begin{itemize}
\item{Spin is a fundamental property of matter and all elementary particles but the Higgs carry spin. Spin cannot be explained by a static picture of the proton, it is caused by the interplay between the intrinsic properties and interactions of quarks and gluons. How does the property of spin of protons and neutrons emerge from quarks and gluons and their underlying interactions? }
\item{Binding leads to a 10$^{-8}$ correction to the mass of an atom, a 10$^{-2}$ correction to the mass of a nucleus, and a factor of 100 correction to the mass of a proton. How does the mass of the nuclear building blocks, the cornerstone of visible matter, emerge from quark-gluon interactions?}
\item{How are the quarks and gluons inside the protons, neutrons and atomic nuclei distributed in both momentum and position space? How can we understand the dynamical origin of these distributions in QCD? What is the relation to confinement?}
\item{How do color-charged quarks and gluons
interact with a nuclear medium? How do the confined color-neutral hadronic states emerge from these quarks and gluons? How do the quark-gluon interactions create nuclear binding?}
\item{How does a dense nuclear environment affect the dynamics of quarks and gluons, their correlations, and their interactions? What happens to the gluon density in nuclei? Does it saturate at high energy, giving rise to gluonic matter or a gluonic phase with universal properties in all nuclei and even in nucleons? Can we see the approach to when the rates of gluon splitting and gluon recombination become equal?}
\end{itemize}

The way in which a nucleon or nucleus reveals itself in an experiment depends on the kinematic regime probed. A dynamic structure of quarks and gluons is revealed when probing nucleons and nuclei at higher energies, or with higher resolutions. Here, the nucleon transforms from a few-body system with its structure dominated by the three valence quarks to a regime where it is increasingly dominated by gluons generated through gluon radiation, as discovered at the former HERA electron—proton collider at DESY~\cite{H1:2015ubc}.
Eventually, the gluon density is predicted to become so large that the gluon radiation is offset by gluon recombination, leading to a saturated state of gluons if both balance each other.

\begin{figure}[!ht]
\begin{center}
\includegraphics[width=0.99\textwidth]{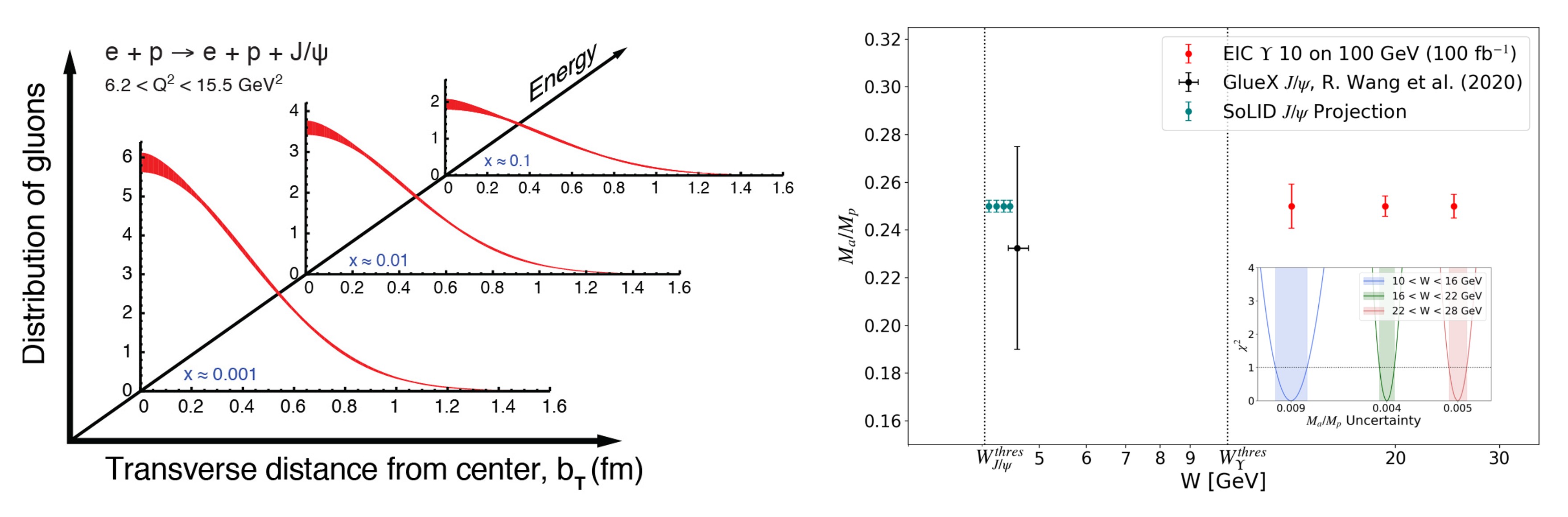}
\caption{
(Left) Projected measurement precision of the distribution of gluons in transverse space inside a proton, obtained from exclusive J/$\psi$ production at the EIC. Projections are shown for three different bins in the fraction of the proton’s momentum carried by the gluons~\cite{Accardi:2012qut,EIC:LRP}.
\quad 
(Right) Projection of the trace anomaly contribution to the proton mass ($M_a/M_p$) with $\Upsilon$ photoproduction on the proton at the EIC in 10 $\times$ 100 GeV electron/proton beam-energy configuration. The insert panel illustrates the minimization used to determine the uncertainty for each data point. The black circles are the results from the analysis of the GlueX J/$\Psi$ data, while the dark green circles correspond the JLab SoLID J/$\Psi$ projections\cite{AbdulKhalek:2021gbh}.}
\label{fig:EICtrans}
\end{center}
\end{figure}

The design parameters
of the EIC are well matched to reveal the quark-gluon structure of matter from the domain where a few valence quarks prevail to the domain where gluons dominate and the non-linear dynamics of the gluon’s self-interactions set in.
The luminosity of the EIC will render precision structure function measurements spanning low-$x$ extending into the valence regime at high-$x$.
The EIC opens up the unique opportunity to go far beyond the present one-dimensional picture of nuclei and nucleons, where the composite nucleon appears as a bunch of fast-moving (anti-)quarks and gluons whose transverse momenta or spatial images are not resolved. It will enable nuclear “femtography” by correlating the information of the quark and gluon longitudinal momentum component with its transverse momentum and spatial distribution inside the nucleon (see Fig.~\ref{fig:EICtrans} (left) for an example). Such femtographic images will provide, for the first time, insight into the QCD dynamics inside hadrons, such as the interplay between sea quarks and gluons, and how this QCD dynamics leads to the known mass (see Fig.~\ref{fig:EICtrans} (right)) and spin-spatial substructure of light hadrons – pions, kaons, protons and neutrons, and its impact in dense nuclear matter. The ultimate goal of the EIC is to reconstruct and constrain experimentally the so-called Wigner functions -- the quantities that encode
tomographic information and constitute a QCD genetic map of nucleons and nuclei.

The versatile EIC will for the first time be able to systematically explore and map out the dynamical system that is the ordinary QCD bound state, triggering a new area of study of the structure of visible matter at its most fundamental level.
The EIC will simultaneously provide promising synergies with the science program at the LHC~\cite{Angeles-Martinez:2015sea,Boussarie:2023izj}.
The EIC is unique and draws large interest all around the world. The EIC Users Group has over 1550 members of which near-half represent 40 non-US countries in six continents.

\vspace{6pt}
\noindent
{\bf Methodology:}
The EIC will study the substructure of protons, neutrons, and atomic nuclei with the most powerful electron microscope, combining versatility, resolving power and intensity. EIC provides the capability of colliding beams of polarized electrons with polarised beams of
protons and light ions.
The EIC accelerator complex is designed to provide high luminosity collisions above 10$^{33}$ and up to 10$^{34}$ cm$^{-2}$ s$^{-1}$ over a wide center-of-mass (CM) energy range of 30-140 GeV. The complex can deliver 70\% or more polarised electron and light-ion beams, and a large range of unpolarised ion beams. The complex comprises 
an existing hadron complex including a modified hadron storage ring (HSR), a new electron storage ring (ESR) with injector, and a new high-luminosity interaction region (IR) including a 25 mrad crossing angle and beam crabbing. Electron cooling is done at hadron injection energies (25 GeV/u) to flatten the hadron beams.

The planned EIC detector (ePIC) combines multiple novel different technologies related to tracking, particle identification, calorimetry, far-forward particle detection and streaming-readout compatible electronics that point the way for future particle physics facilities in Europe.  

The HSR magnet and vacuum systems from the Relativistic Heavy Ion Collider (RHIC) will be reused, though the average beam current will be tripled as compared to the 1 A of RHIC operations.
There is active collaboration with INFN to characterize the secondary electron yield for electron cloud suppression.

The ESR average beam current (up to 2.5 A) and bunch parameters
are similar to those envisioned for FCC-ee. There is active collaboration between EIC and CERN on diagnostics technologies for beams in this parameter range, including beam position monitoring and polarimetry.

Hadron beam crabbing is required for both EIC and HL-LHC, albeit with much larger crab crossing angles at EIC of 25 mrad. A collaboration of EIC labs (BNL, Jefferson Lab) and CERN demonstrated and characterized hadron beam crabbing successfully in the SPS in 2018. 

The EIC design requires new NbTi superconducting magnets for the interaction region final focus and spectrometer bends, and electron beam spin rotator solenoids. Both traditional collared and direct wind magnets are envisioned, 
driven by space constraints and large aperture and field requirements.

Electron cooling to create “flat” hadron bunches is planned at hadron injection energy (25 GeV/u). This is based on the proven technology used for low energy RHIC cooling scaled to higher energy. High-energy strong hadron cooling is a possible future upgrade to double the integrated luminosities.

\vspace{6pt}
\noindent
{\bf Readiness:} The scientific foundation for the Electron-Ion Collider (EIC) was built over two decades.
The EIC was
supported by the United States Nuclear Science Advisory Committee Long-Range Plans of 2002 and 2007, and a top recommendation of the 2015 and recent 2023 plan. The EIC addresses unique and compelling science as evidenced by a 2018 consensus study report of the National Academies of Science, Engineering and Medicine. 
EIC construction readiness is demonstrated by the following major milestones: 
\begin{itemize}
\setlength\itemsep{-0pt}
\item{The EIC received Critical Decision 0 (CD-0) mission-need in December 2019 and CD-1 in June 2021.}
\item{The EIC received \$138M of Inflation Reduction Act funding in September 2022.}
\item{The electron-Proton Ion Collider (ePIC) detector collaboration was formed in July 2022. In 2024 ePIC was approved as a
CERN recognized experiment.}
\item{EIC Resource Review Board meetings were started in 2023, see https://www.bnl.gov/eic-rrbmeeting/.}
\item{The State of New York awarded \$100M for the construction of EIC support buildings in February 2024.}
\item{EIC received Critical Decision 3A (CD-3A) for the start of long lead procurements in March 2024, and is anticipated to receive shortly a further Critical Decision 3B (CD-3B).}
\item{The R\&D performed to date confirms the technical feasibility of the facility.}
\item{Critical Decision 2, Performance Measurement Baseline, and Critical Decision 3, Approve Start of Construction, are on track for DOE approval in 2026.}
\end{itemize}
Formal agreements for international in-kind contributions, notably those confirmed from Canada, Italy and the UK, are in development. Statements of Interest were signed by the DOE and French agencies. Proposals for possible in-kind contributions from Japan and Korea are in final stages of review.

\vspace{6pt}
\noindent
{\bf Challenges:}
Hadron crabbing still faces considerable challenges to achieve the low crab cavity phase noise tolerances, crab bump closure tolerances, coupling compensation, and multipole considerations for beam dynamics. Collaboration with CERN 
and worldwide experts is ongoing to solve these common problems for both EIC and HL-LHC. Similarly, the crab cavity fabrication and engineering are challenging to achieve the required large EIC crab crossing angle, with complicated crab cavity shapes that require careful 3D assembly processes. EIC collaborations with CERN and ASTEC/Cockroft Institute in the UK are
developing
international crab cavity and cryomodule production capabilities for future colliders including EIC.

The EIC design requires new NbTi superconducting magnets for the interaction region (IR) final focus and bends, and electron beam spin rotator solenoids. Several of these are challenging to fulfill the aperture requirements of the IR. The spin rotator solenoids are challenging due to their high fields (upwards of 10T), transverse space constraints, and shielding requirements that preclude passive shielding. Areas of collaboration include design and construction of the IR magnets (various institutions), the superconducting cable production (CERN), and spin rotator solenoid design/production (CEA Saclay).
The EIC will require the further realisation of polarized ion sources and transport and optimization of polarized electron and proton/light-ion beams, and accompanying polarimeters.

A future upgrade of strong hadron cooling by electrons to double the integrated luminosities requires a high-current (upwards of 100 mA) ERL with the average beam power approaching 10 MW. This is beyond the state-of-the-art, but comparable to and synergistic with European ERL activities on facilities such as bERLinPro (HZB) and PERLE (IJCLab). Jefferson Lab presently holds the record for demonstrated ERL beam power and is an active collaborator on PERLE as a test bed for a possible LHeC electron-hadron collider at CERN. All envisioned high-power ERLs also require state-of-the-art developments for high-current superconducting RF (SRF) cavities and couplers and understanding of beam dynamics limitations from SRF and magnet field quality.

\subsection{Fixed Target DIS at Jefferson Laboratory}
\label{jlab}

\noindent
{\bf Science Potential:} The Jefferson Lab fixed-target program offers very high luminosities \mbox{(10$^{35-39}$ cm$^{-2}$ s$^{-1}$)} 
at electron scattering energies of up to about 11 GeV~\cite{Dudek:2012vr}. This allows precision measurements of both spin-averaged and spin-dependent structure functions in the Bjorken-$x$ range of 0.1-0.8, and access to 3D transverse spatial and momentum imaging in the valence quark region. A JLab upgrade can offer critical insights for precision studies of partonic structure filling the gap in kinematics of the combined scientific program of the present JLab and the foreseen EIC. With its enhanced energy range, the JLab 22 GeV upgrade~\cite{Accardi:2023chb} will allow:

\begin{itemize}

\item {Precision measurements of the nucleon light sea in the intermediate to high-$x$ range, from $x \sim$ 0.05 up to $x$ = 0.9,
which can help validate novel theoretical predictions for the intrinsic sea components in the nucleon wave function and support Beyond Standard Model searches at colliders~\cite{Ablat:2024muy}.}

\item {Play a fundamental role in the EIC era in the realm of precision separation measurements between the longitudinal ($\sigma_L$) and transverse ($\sigma_T$) photon contributions to the cross section, which are critical for high-precision studies of both semi-inclusive and exclusive processes.}

\item {Precision determination of the helicity structure of the nucleon at large $x$ and of the strong coupling in the region bridging confinement to asymptotic freedom. Together with projected EIC data this can halve the uncertainty of the most precise world data extraction of $\Delta \alpha/\alpha$ to $\sim$0.6\%.}

\item {Unique opportunities to explore the internal structure of mesons from intermediate to high-$x$~\cite{Barry:2018ort,Cao:2021aci,Alexandrou:2024zvn}.}

\end{itemize}

\noindent
\begin{minipage}{1.0\textwidth}
\begin{minipage}{0.40\textwidth}

\noindent
{\bf Methodology:} The proposed energy upgrade of the \mbox{CEBAF} accelerator up to 22 GeV can be achieved by increasing the number of recirculations through the accelerating cavities within the existing tunnel footprint. Encouraged by the recent success of the CBETA project (Cornell Brookhaven Electron Test Accelerator), the highest-energy arcs would be replaced with Fixed Field Alternating Gradient (FFA) arcs. The new
arcs configured with an FFA lattice would support simultaneous transport of 6 passes with energies spanning a factor of two. This wide energy bandwidth could be achieved using the non-scaling FFA principle implemented with Halbach-style permanent magnets. This novel magnet technology saves energy and lowers operating costs. Such a scheme would nearly double the energy while using the existing CEBAF SRF cavity system.

\end{minipage}
\hspace{0.05\linewidth}
\begin{minipage}{0.50\linewidth}
\begin{figure}[H]
\null\vspace{-0.0cm}
    \includegraphics[width=0.85\textwidth]{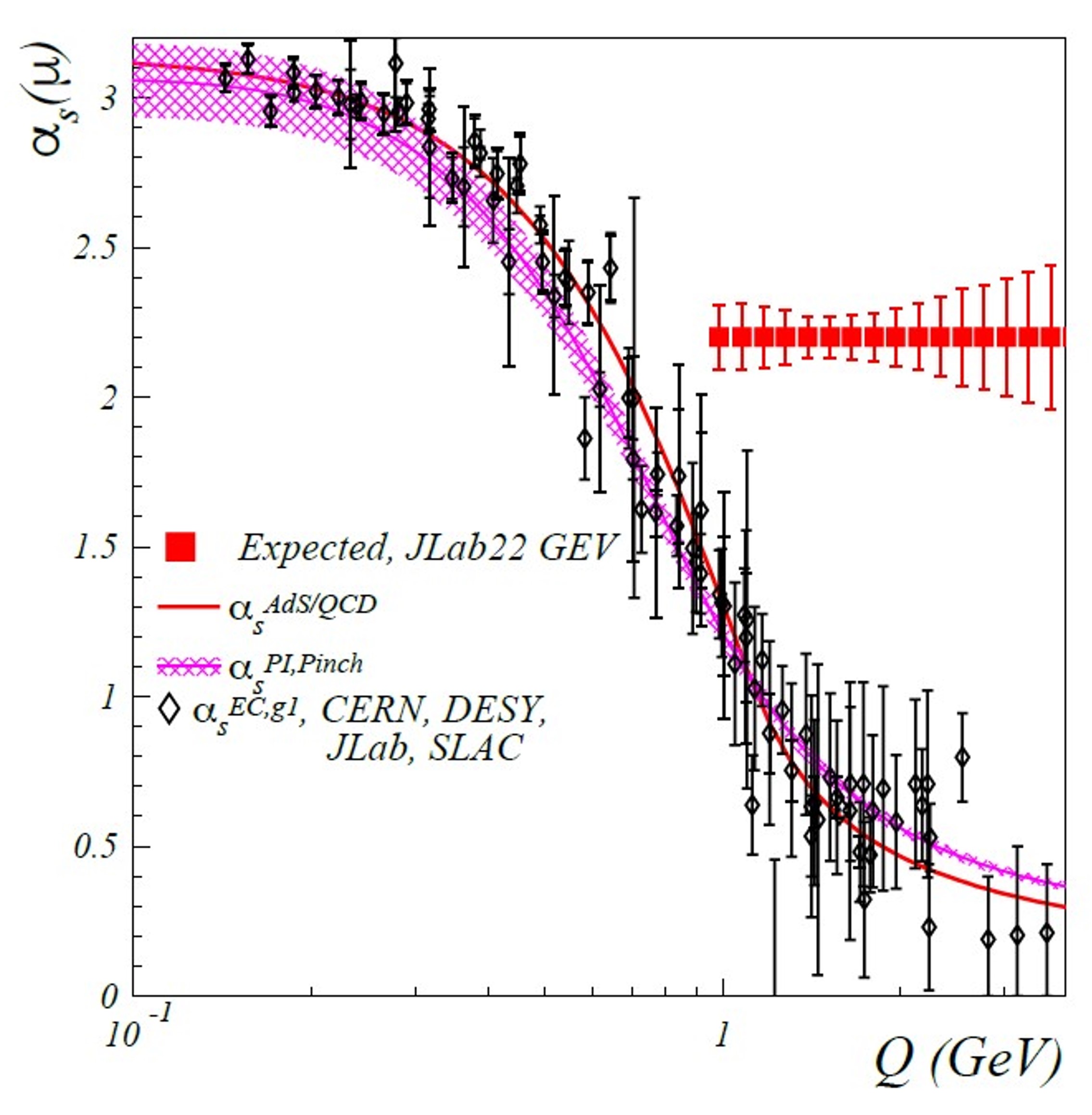}
    \null\vspace{-20pt}
    \caption{
    Expected accuracy on mapping $\alpha_s(Q^2)$ (squares) from the Bjorken Sum Rule utilizing JLab at 22 GeV~\cite{Accardi:2023chb}, as compared to world data and predictions. Projected data are shown at an arbitrary value.}
    \end{figure}
\end{minipage}
\hfill
\end{minipage}
\vspace{0pt}

\vspace{6pt}
\noindent
{\bf Readiness:} It is important to note that demonstrating the ability to scale FFA technology up in energy is critical to the above planning and science. At this point, Jefferson Lab proposes to work on this critical step in the near future, with a longer timeline envisioned for the full upgrade. The readiness for science operations is projected to be in the
late 2030s.

\vspace{6pt}
\noindent
{\bf Challenges:} Staying within the \mbox{CEBAF} footprint while transporting high energy beams calls for special mitigation of synchrotron radiation effects. Methods to preserve acceptable emittance and energy spread are under study. A further challenge is the method of beam extraction from the multi-pass higher energy beams. 

\subsection{Neutrino DIS at CERN}
\label{fpf}

\noindent
{\bf Science Potential:}
Forward particle production in proton-proton collisions at the LHC produce a large flux of the highest-energy (TeV-scale) neutrinos ever produced in a laboratory~\cite{Buonocore:2023kna}.
The recent groundbreaking observation of neutrinos by the FASER~\cite{FASER:2023zcr} and SND@LHC~\cite{SNDLHC:2023pun} far-forward LHC experiments, together with the first differential measurements of the neutrino interaction cross-section at TeV energies by FASER$\nu$~\cite{FASER:2024ref,FASER:2024hoe}, herald the beginning of the collider neutrino era, with neutrino-induced DIS as core underlying process.\\[-0.3cm]

\begin{figure}[t]
\centering
\includegraphics[width=0.99\textwidth]{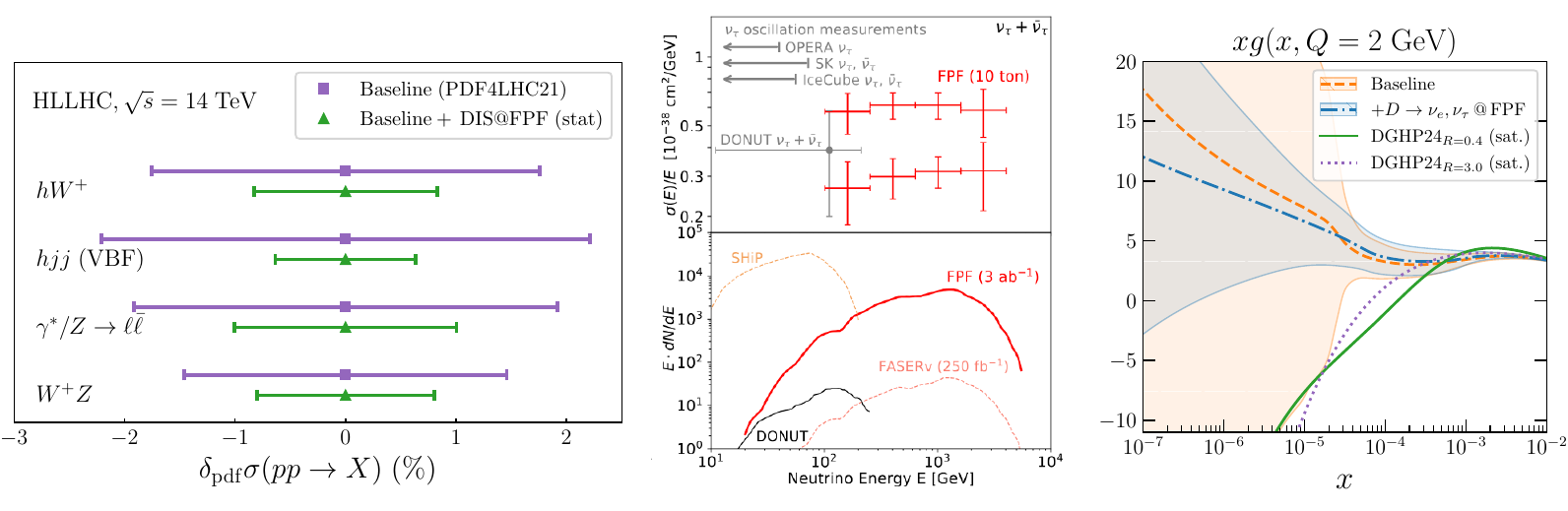}
\vspace{-0.3cm}
\caption{
Representative applications of LHC neutrino DIS: reduced PDF errors for HL-LHC cross-sections~\cite{Cruz-Martinez:2023sdv} (left),
fingerprinting tau neutrinos~\cite{Adhikary:2024nlv} (middle), and probing the small-$x$ gluon~\cite{Rojo:2024tho} (right).
}
\label{fig:FPF-HLLHC}
\end{figure}

\noindent
Going beyond these Run-3 pathfinder experiments, exploiting neutrino DIS in the HL-LHC era would sharpen our understanding of proton and nuclear structure, enhance the reach of new particle searches at ATLAS and CMS, and provide key input for the modelling of astroparticle experiments such as IceCube and Auger.
The novel opportunities for particle, hadronic, and astroparticle physics of neutrino DIS at the LHC have been summarised in~\cite{Anchordoqui:2021ghd,Feng:2022inv,Adhikary:2024nlv}, and are illustrated in Fig.~\ref{fig:FPF-HLLHC} with three examples: the reduction of PDF errors for HL-LHC cross-
sections enables by neutrino DIS, fingerprinting tau neutrino interactions, and probing the small-$x$ gluon PDF from forward charm production.
The most direct way to realize this unique potential is via a dedicated Forward Physics Facility (FPF)~\cite{Anchordoqui:2021ghd,Feng:2022inv,Adhikary:2024nlv}, a new underground cavern at CERN to house a suite of far-forward experiments during the HL-LHC.
The FPF is projected to detect $\mathcal{O}\left( 10^6\right)$ electron and muon neutrinos and $\mathcal{O}\left( 10^4\right)$ tau neutrinos during the full HL-LHC operations ($\mathcal{L}_{\rm pp}=3$ ab$^{-1}$),
enabling precision probes of neutrino properties and interacting for all 3 flavors and a rich program of TeV-energy neutrino DIS~\cite{Cruz-Martinez:2023sdv,Candido:2023utz}. 
Furthermore, FASER has been approved to operate during Run-4, and a new AdvSND experiment~\cite{Paggi:2024fae} upgrading SND@LHC has also been proposed.\\[-0.4cm]

Recent studies have highlighted the potential of LHC neutrino DIS  to measure the gluon PDF down to very small-$x$~\cite{Rojo:2024tho} (right panel of Fig.~\ref{fig:FPF-HLLHC}), constrain forward hadron production~\cite{Kling:2023tgr}, test non-linear QCD dynamics~\cite{Bhattacharya:2023zei}, identify a possible intrinsic charm contribution in the proton~\cite{Ball:2022qks,Guzzi:2022rca}, map large-$x$ nucleon structure and hence improve predictions for Higgs and gauge boson production at the HL-LHC~\cite{Cruz-Martinez:2023sdv} (left panel of Fig.~\ref{fig:FPF-HLLHC}), break degeneracies between possible BSM signals and QCD effects in high-$p_T$ tails at the HL-LHC~\cite{Hammou:2023heg,Hammou:2024xuj,Fiaschi:2022wgl,Fiaschi:2021sin}, test lepton flavour universality in the neutrino sector, and tune generators for neutrino astrophysics~\cite{vanBeekveld:2024ziz,FerrarioRavasio:2024kem,Buonocore:2024pdv}, among many other applications.\\[-0.4cm]

Looking to the more distant future, 
FPF linked to
beams produced in proton-proton collisions at the 
FCC-hh~\cite{MammenAbraham:2024gun} 
(see also section~\ref{fcceh})
would be the 
ultimate neutrino DIS experiment,
further 
constraining the unpolarised, nuclear, 
and spin structure of the nucleon in extreme regions.

\vspace{6pt}
\noindent
{\bf Methodology:}
At Run-3, reconstructing neutrino DIS is accessible via FASER$\nu$ and SND@LHC emulsion detectors, with complementary sensitivity provided via the FASER electronic components.
Neutrino DIS at the FPF will be possible through two independent detectors: FASER$\nu$2, an on-axis emulsion detector with unparalleled spatial resolution, and FLArE, a 10-ton-scale noble liquid fine-grained time projection chamber with high kinematic resolution and wide dynamic range.
Different options for a new neutrino detector at FASER during Run-4 are currently being discussed.
\\[-0.4cm]

\vspace{6pt}
\noindent
{\bf Readiness:}
An extensive site selection study has been conducted by the CERN Civil Engineering group, and a suitable candidate site for the FPF has been identified.
This site is shielded from the ATLAS IP by over 200~m of concrete and rock and exhibit with geological conditions enabling immediate construction.
Vibration, radiation, and safety studies show that the FPF can be constructed independently of the LHC without interfering with its operations. 
All of the FPF experiments are relatively small, low cost, require limited R\&D, and can be constructed in a timely way. 
A Class 4 cost estimate for the FPF is 35 MCHF for the construction of the new shaft and cavern, with neutrino experiments costing from roughly 15 MCHF for FASER$\nu$2 to 30~MCHF for FLArE.
 A possible timeline is for the FPF to be built during Long Shutdown~3, the support services and experiments to be installed starting in 2029, and the experiments to begin taking data during Run~4. 
\\[-0.4cm]

\vspace{6pt}
\noindent
{\bf Challenges:}
To fully exploit the far-forward physics opportunities, many of which may disappear for several decades if not explored in the 2030s, the FPF and its experiments should be ready for physics data-taking in the HL-LHC era as early as possible in Run 4. 

\subsection{The Large Hadron electron Collider}
\label{lhec}

\begin{figure}[tb]
\centering
\includegraphics[width=0.48\textwidth]
{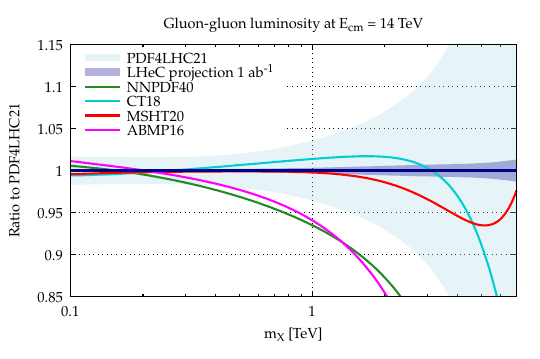}
\hfil
\includegraphics[width=0.48\textwidth]{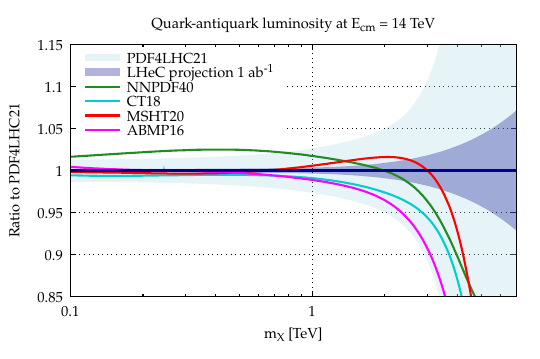}
\caption{Impact of LHeC on the precision of the 
gluon-gluon and quark-antiquark 
luminosities at the \mbox{HL-LHC} as a function of parton-parton 
invariant mass $m_X$, compared
with recent global fits.
Results from the NNPDF \cite{NNPDF:2021njg}, 
CT \cite{Hou:2019efy}, 
MSHT \cite{Bailey:2020ooq}
and ABMP \cite{Alekhin:2017kpj}
groups
based on fits to current data are compared with LHeC projections,
all normalised to PDF4LHC21 \cite{PDF4LHCWorkingGroup:2022cjn}. 
Uncertainty bands are shown for PDF4LHC21 and LHeC. 
Full details in \cite{lhec:whitepaper}.
}
\label{fig:LHeC-partons}
\end{figure}

The Large Hadron electron Collider 
(LHeC)~\cite{Dainton:2006wd,AbelleiraFernandez:2012cc,LHeC:2020van,Andre:2022xeh,lhec:whitepaper}
is the most readily obtainable step forward in 
centre-of-mass energy for a DIS collider, entering the TeV
regime for the first time, whilst also offering 
luminosities of order 1000 times larger than that of HERA. 
By adding an electron beam, nominally of 50 GeV energy, 
based on an energy-recovery linear accelerator (ERL) to the 
existing CERN proton and 
heavy ion accelerator complex, it has the potential to 
significantly extend the science programme of the LHC, 
ensure continuity of energy-frontier collisions in the 2040s and 
bridge towards the next flagship collider in terms of technology
development and scientific scope. 
Its modest requirements in new tunnel length and
exploitation of existing infrastructure makes it affordable 
within the existing CERN budget, whilst its 
use of superconducting 
RF and energy recovery make it a testing ground for 
sustainable future accelerator techniques.

\vspace{6pt}
\noindent
{\bf Science Potential:}
The LHeC has a very broad physics scope. On the one hand, 
it will operate as a scattering experiment to resolve 
hadronic structure and explore QCD dynamics in new kinematic 
regions for both protons and heavy nuclei. On the other hand
it has much in common with the general purpose detectors 
at the
LHC, having its own standalone Higgs, top, electroweak and 
BSM search programmes. 
A brief discussion is given below, restricted to 
the key aspects of proton parton density determinations and 
Higgs coupling sensitivities. 
Much more detailed information, including studies
spanning the full range of relevant physics
topics, is available in the recent updated Conceptual Design 
Report \cite{LHeC:2020van} and White Paper \cite{lhec:whitepaper}.

As a machine for measuring hadron structure, the LHeC has the
capability to revolutionise our understanding of the collinear
parton densities of protons, as well as the modifications that
take place in nuclei. This includes a full flavour decomposition,
a new level of precision at large $x$ and a first exploration
of the very low $x$ region extending to $x \sim 10^{-6}$.
In this low $x$ region, it is widely expected that new 
strong interaction dynamics are required in order to tame the
growth of the gluon density, a fundamental boundary of our current 
understanding. 

The expected LHeC PDF precision is propagated to the
corresponding uncertainties on the parton-level initial state 
in LHC proton-proton collisions (the so-called `parton 
luminosity') in figure~\ref{fig:LHeC-partons}. 
There is a 
significant improvement over current global fits throughout the
accessible range of produced particle 
or system invariant masses $M_X$. Applied to
current and future LHC analysis at large $M_X$, this
leads to improved sensitivity for a 
wide range of searches for new physics. 
In the intermediate $M_X$
region, the LHeC dramatically decreases the (often 
dominant) systematic uncertainties 
due to PDF inputs to many LHC 
measurements of strong interactions
and electroweak parameters. Examples include the $W$ boson mass
measurement, where the PDF uncertainties are expected 
in the ATLAS context to be 
reducible from around $8 \ {\rm MeV}$ now 
to $2 \ {\rm MeV}$ \cite{ATLAS:2018qzr}, 
and 
$\sin^2 \theta_W$, 
which can be constrained to a precision of
0.03\% using LHeC data. 
Similarly, the LHeC in $eA$
mode provides a new level of 
understanding of nuclear PDFs in the
kinematic range relevant to 
fully exploit heavy ion collisions at the 
LHC and at future colliders. 

\begin{figure}[tb]
\centering
\includegraphics[width=0.49\textwidth]{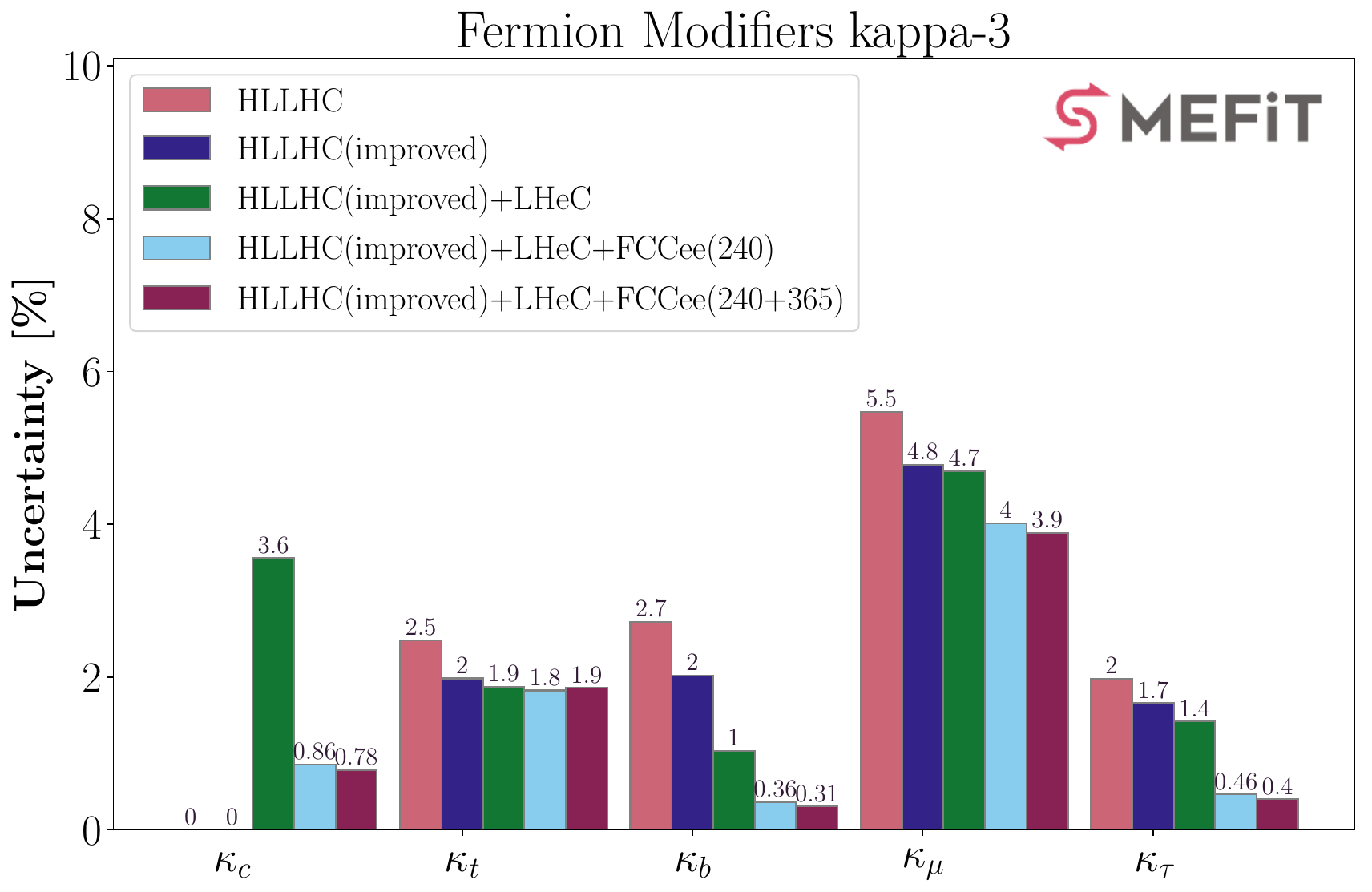}\hskip -0.1cm
\includegraphics[width=0.49\textwidth]{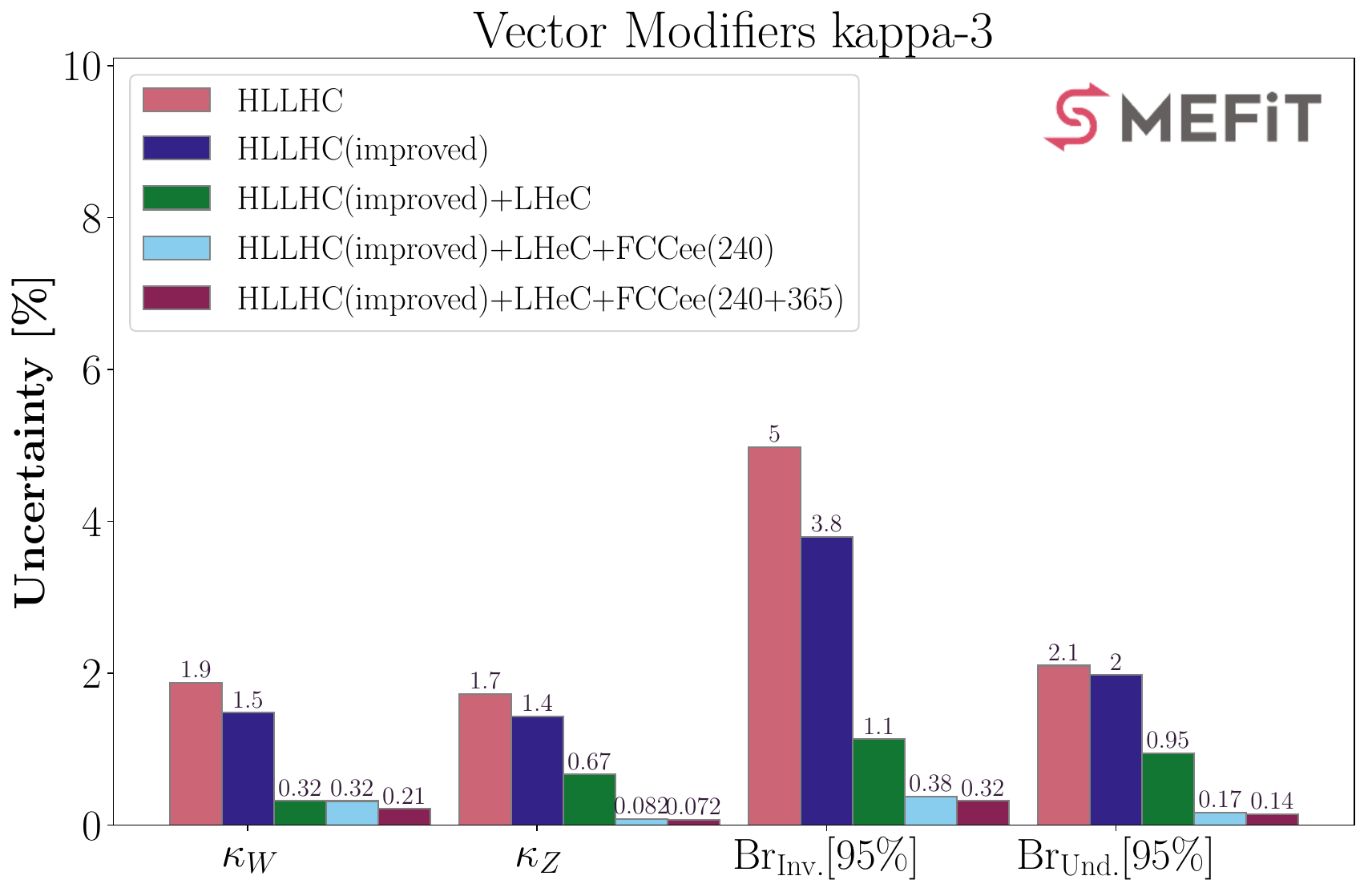}
\caption{
Comparison~\cite{lhec:whitepaper}
of the expected precision on 
example Higgs couplings in the kappa-3 framework~\cite{deBlas:2019rxi} at the \mbox{HL-LHC},
the HL-LHC with improved PDFs from LHeC,
the HL-LHC in combination with the
LHeC, and the HL-LHC 
in combination with the FCC-ee.
Left: fermion coupling modifiers. 
Right: vector coupling modifiers,
including branching ratios  
$BR_{\rm inv}$ to invisible states and 
$BR_{\rm unt}$
to visible states such as $gg$ that are 
not directly measurable due to backgrounds.
The impact of the LHeC over HL-LHC is substantial
in all cases and remains an important ingredient
even with the addition of FCC-ee constraints. 
The 
results were obtained with the {\sc\small SMEFiT} framework~\cite{Giani:2023gfq,Celada:2024mcf,terHoeve:2023pvs} using the inputs from~\cite{deBlas:2019rxi}.
}
\label{fig:LHeC-Higgs}
\end{figure}

An illustration of the potential of the LHeC's 
Higgs programme 
is shown in Fig.~\ref{fig:LHeC-Higgs},
for the scenario where the LHeC bridges
between the HL-LHC and the FCC-ee, in the
kappa-3 framework 
\cite{Giani:2023gfq,terHoeve:2023pvs, Celada:2024mcf}.
The potential improvements due to the LHeC are shown 
both through its impact on HL-LHC results via 
PDF determinations and through the 
direct addition of LHeC data. 
Various other scenarios
and couplings are considered in \cite{lhec:whitepaper}.
The total LHeC 
cross section for Higgs production is at the level
of $0.2 \ {\rm pb}$ assuming 80\% lepton polarisation,
leading to yields at the level of $2 \times 10^5$ for
$1 \ {\rm ab^{-1}}$. 
The dominant charged-current production mechanism,
$ep \rightarrow \nu HX$ with a $WW \rightarrow H$
subprocess, naturally
leads to very 
strong sensitivity to the Higgs coupling to the $W$
boson, which remains the dominant 
constraint in Fig.~\ref{fig:LHeC-Higgs}
until FCC-ee running at the
$t \bar{t}$ threshold. 
The sub-leading neutral current $ZZ \rightarrow H$
channel is easily distinguished experimentally and also 
improves substantially on the expected precision from the HL-LHC. Meanwhile,
the relatively clean events and lack of pile-up offer
an improved window on
the beauty channel and promise what 
would likely be the
first $5 \sigma$ observation of the Higgs decaying to charm 
quarks. Significant improvements from LHeC over HL-LHC are also
visible in most of the other channels. 
Overall, 
the improvement in precision on Higgs couplings when adding the 
LHeC to the HL-LHC is at a similar level to the
gain when moving from the LHC to the HL-LHC. 
There is thus considerable complementarity between the
sensitivities of the LHC and the LHeC. 
The improved sensitivities from LHeC
are also complementary to those obtainable
with FCC-hh \cite{lhec:whitepaper}

\vspace{6pt}
\noindent
{\bf Methodology:}
The LHeC can provide electron-proton and electron-ion collisions
with very high energies, $\sqrt{s}$ = 1300 GeV, and high luminosity, of order $10^{34}$~cm$^{-2}$s$^{-1}$. 
Two superconducting linacs
accelerate the electrons in a 3-turn racetrack configuration 
up to 50 GeV,
before recovering the bulk of the energy used by decelerating
through the same RF cavities with the phase inverted. 
The wall-plug power has been constrained to 100 MW 
and the tunnel length is 5.4~km for the baseline 50 GeV design.  
The LHeC would occupy one of the existing LHC
Interaction Points (IP); civil engineering options have been 
considered primarily for IP2. 
Polarization of the electron beams is anticipated.

Operating as a standalone facility beyond the end of the \mbox{HL-LHC}, 
the estimated integrated luminosity is $180 \ {\rm fb^{-1}}$ 
per year, allowing timely accumulation of 
the target $1 \ {\rm ab^{-1}}$. 
Concurrent operation with 
other experiments has also been studied and would be 
possible, albeit at reduced luminosities 
of around $50 \ {\rm fb^{-1}}$ per year, if standard
\mbox{HL-LHC} operations continue for long enough for 
the LHeC to be realised.

The LHeC detector has a classic collider-experiment structure, 
with an asymmetric design accounting for 
the differences between the electron and proton/ion energies. 
Dipoles are used to bring the electron beam into head-on collisions
with the proton. These, combined with a central solenoid,
are inserted between the electromagnetic and hadronic calorimeters. 
An IR design exists that accommodates a 
spectator proton beam, as would be required for 
concurrent LHeC and LHC operations. 

The central LHeC detector must be complemented by extensive
beam-line instrumentation. Forward proton and neutron taggers 
enable a programme of diffractive physics, while the backward region includes near-axis photon and electron detectors
to determine the luminosity from Bethe-Heitler 
scattering and study quasi-real photoproduction. 

\vspace{6pt}
\noindent
{\bf Readiness:}
The LHeC is targeted by default for operation in the 2040s,
coming online soon after the end of the \mbox{HL-LHC}. It could,
however, be ready in the late 2030s 
to allow concurrent running
with other LHC experiments. 
The ERL-based electron design 
has the advantage of being realisable, to considerable extent, 
independently of the LHC operations. The ERL arrangement is located inside the LHC ring but outside the tunnel to minimize interference.  

Key technology developments include large low-mass high-resolution tracking detectors, in synergy with the LHC luminosity upgrade, but with largely reduced radiation and pile-up levels. The current
designs under development for the EIC make an excellent starting
point. The same applies 
to the forward and backward beam-line instrumentation, which are
key to much of the EIC physics programme and where technology
transfers to the LHeC look attractive.

\newpage\goodbreak
\vspace{6pt}
\noindent
{\bf Challenges:}
Demonstration of the multi-pass and multi-GeV ERL technology 
at high currents
is the main challenge and is the subject of significant R\&D, motivated by multiple potential applications  
in future energy-frontier electron 
acceleration \cite{Telnov:2023vlz}.
The PERLE \cite{Angal-Kalinin:2017iup} facility at IJCLab is 
positioned as a prototype for the LHeC, with two 
$80 \ {\rm MeV}$ ERL linacs based on 
$802 \ {\rm MHz}$ superconducting cavities already successfully built at JLab. The current PERLE programme aims for a single
pass machine to be operational by 2028. With sufficient
funding, a 3 turn version could be in place for 2030. 

Further ongoing R\&D is related to the high-current (polarised) electron source utilised in the linac-ring design, and the ability of the SRF cavities to operate with high average and peak beam currents, and the accompanying cryomodule design to contain the high beam power.
The latter questions have synergy with FCC needs. 

\subsection{The Future Circular Collider in electron-hadron mode}
\label{fcceh}

The 90 km circumference tunnel under consideration for the Future Circular Collider (FCC)
at CERN 
ultimately enables collisions between 
electrons (e),
hadrons (h), 
or the combination of 
the two \cite{FCC:2018byv,FCC:2018vvp}. 
The FCC-eh design
exploits the ERL-based linac-ring design of the LHeC 
to enable electron-proton collisions at
$\sqrt{s}$ = 3.5 TeV, a factor of 3 beyond LHeC, 
with a comparable 
luminosity of 
$\sim 1.5\times 10^{34} \ {\rm cm^{-2}s^{-1}}$. 
In two decades of concurrent operation with FCC-hh, an integrated luminosity of 2-3 ab$^{-1}$ may be collected.

\vspace{6pt}
\noindent
{\bf Science Potential:}
The unprecedented 
energy and luminosity achievable at FCC-eh would
enable sub-percent precision Higgs 
coupling measurements, as well as electroweak and
top quark measurements and searches for new physics
including extra neutral and charged gauge bosons,
prompt and long-lived new scalars, heavy 
leptons, dark photons and axions. The sensitivity to
proton structure extends to $x$ values as low as $10^{-7}$
with correspondingly enhanced sensitivity to novel QCD
dynamics and providing the only realistic pathway to a
well-understood initial state for the FCC-hh.

\vspace{6pt}
\noindent
{\bf Methodology:}
The FCC-eh and FCC-hh are designed to operate concurrently,
with the current design placing FCC-eh at
point L on the FCC ring, not far from CERN.
The 60 GeV electron beam from the ERL would hit head on with 
one of the proton beams of the FCC while the other 
proton beam bypasses the interaction region. 
Scaling up the LHeC solutions to account for
higher energies leads to a workable first-pass 
detector design. 

\vspace{6pt}
\noindent
{\bf Readiness:}
The technology requirements of the FCC-eh interaction 
region and detector in their current designs are 
not significantly different from those of the LHeC. 
In particular, if the ERL energy and current scalabilities
are demonstrated for the LHeC, they are equally applicable
for FCC-eh. 

\vspace{6pt}
\noindent
{\bf Challenges:}
As with the FCC-hh, the main obstacle to 
realising FCC-eh 
is high-field dipole development. The large asymmetry between the 
proton and lepton beam energies is also challenging.

\linespread{0.94}\selectfont

\subsection{DIS at CERN with Plasma-wakefield acceleration}
\label{vheep}
\null \vspace{-12pt}
An electron and positron acceleration scheme based on 
{\bf proton-driven plasma wakefield acceleration} (PDPWA) would in principle allow for much higher energy collisions than pure ring-based schemes. In the previous EPPSU, we discussed AWAKE-like schemes of modulating long proton bunches~\cite{AWAKE:2022aeo} and using the modulated proton bunch to drive the plasma wakefield. Here, we focus on a recent concept~\cite{Farmer:2024xqm} using short proton bunches where additionally  high repetition rates could 
be achieved, leading to much higher luminosities. The accelerated electrons and positrons would be collided with LHC protons and ions to produce very high energy eP and eA collisions.  An initial implementation would use the SPS tunnel, and we assume that the SPS and its pre-injectors have been retrofitted with a
Fixed-Field Alternating Gradient (FFA) Accelerator
as described in~\cite{ref:ALIVE}.  Such an FFA scheme is expected to yield short proton bunches of $10^{11}$ protons/bunch and up to $E_p \simeq 550$~GeV per proton at a rate of $\simeq 20$~kHz.  With protons of this energy, we expect that  bunches of $2\cdot 10^{10}$ electrons or positrons can be accelerated up to $E_e\simeq 250$~GeV in a plasma section, yielding $E_{\rm cm}\simeq 2650$~GeV for collisions with LHC protons. The emittance of the electron and positron (to be confirmed) bunches will be adequate such as to allow a match of the bunch transverse dimensions to the proton bunch size at the IP.  Using the proton bunch parameters from the LHeC studies, a luminosity of $10^{32}$~cm$^{-2}$s$^{-1}$ will be possible. The kinematic range that will become available for DIS experiments using the SPS as proton drive beam will be extended by approximately two orders of magnitude relative to HERA.  If eventually also the LHC tunnel can be used to house the proton driver accelerator, considerably higher final lepton energies will be available, and CoM energies of $E_{\rm cm}=9000$~GeV can be reached~\cite{Caldwell:2016cmw}.

\goodbreak
\vspace{6pt}
\noindent
{\bf Readiness:}
The SPS and its pre-accelerators would need to be retrofitted with the FFA scheme and some limited tunneling would be needed for the plasma sections.  Presumably an existing LHC hall would be available for the experimental detector.  This development would happen after the HL-LHC running and would presumably take a few years.  Earliest collisions would then be in the mid-to-late 2040's. 

\vspace{6pt}
\noindent
{\bf Challenges:}
The FFA scheme is still in its conceptual design phase and still requires a detailed evaluation.  Reaching the short ($\sigma_z=150\; \mu$m) proton bunches required for strong plasma acceleration needs to be demonstrated.  For the plasma acceleration, the necessary plasma channels need to be developed and acceleration of positrons needs to be solved if positrons are to be used.

\vspace{6pt}
\noindent
{\bf Benefits:}
The use of the PDPWA scheme may be compatible with providing a full energy injector for the FCC, so that this investment could have multiple uses.  A further use would be to a linear $e^+e^-$ collider at center-of-mass energies up to $550$~GeV using the SPS FFA.  Once the technology is demonstrated, then the LHC could also be retrofitted with the FFA scheme, allowing for higher energy $e^+e^-$ collisions.

\section{Summary}
\label{summary}
The DIS framework has historically proved to be, continues now, and will remain 
a unique 
tool with widespread applications in our field.
As it explores new kinematic domains and levels of precision, 
it continually brings
deeper insights into the nature of the enigmatic QCD theory 
and facilitates discoveries, 
extending to the Higgs sector and 
Beyond the Standard Model scenarios.

The DIS and related topics document submitted to the 
2018 European Particle Physics Strategy Update stated: ``Deep Inelastic Scattering provides an excellent framework to make a decisive and transformational leap in our understanding of the strong interaction, while also providing discovery potential via precision measurements and sensitivity to the direct production of new particles.'' and went on to say
``The new lepton-hadron facilities … offer dramatically new energy and luminosity reach in the DIS context, as well as a full set of polarisations and a vast range of heavy ions in addition to proton beams, representing an exciting and diverse programme for the medium and long-term future.'' 
These conclusions are even more true now than they were in 2018,
with enormous progress having been made in the intervening period.
Currently fixed target DIS facilities
are exploring new territory, including the Jefferson Lab CEBAF and 
neutrino DIS based on the LHC, whilst 
a US-based Electron Ion Collider is progressing well
towards realisation. 

The foundations for the medium-term future are 
thus set, with DIS facilities both present and on the horizon, promising an exciting and diverse scientific programme founded on precision and discovery in the 
fundamental structure and dynamics of matter.
The essential next step,
which can be uniquely addressed in Europe,
is to add an energy frontier facility.
This brings the 
key additional features of complementary Higgs, top and
electroweak sensitivity
to the HL-LHC and enables HL-LHC discovery through improvements
in PDF precision. In the LHeC, we have a well advanced proposed
facility that can fulfil these ambitions, whilst also
building
a bridge between the HL-LHC and the next flagship European
project in terms of time-schedule, technology development, and
physics exploitation opportunities. 

Looking further into the long-term future, 
extending the LHeC concept to the Future Circular Collider and
applying Plasma Wakefield acceleration to DIS offer pathways
to continue exploring the unknown at 
unprecedented lepton-hadron centre-of-mass energies and parton 
densities.

\newpage

\bibliographystyle{utphys}
\bibliography{refs2024}

\end{document}